\documentclass[12pt]{article}

\usepackage{epsfig}

\newcommand{\bmp}{\mbox{\boldmath$p$}}
\newcommand{\bmq}{\mbox{\boldmath$q$}}

\begin{document}

\title{Prescriptions for the scaling variable of the nucleon structure
function in nuclei}

\author{V. I. Ryazanov, B. L. Birbrair and M. G. Ryskin\\
Petersburg Nuclear Physics Institute\\ Gatchina, St. Petersburg 188300,
Russia } \date{} \maketitle

\begin{abstract}
We tested several choices of the in-medium value of the Bjorken scaling
variable assuming the nucleon structure function in nucleus to be the
same as that of free nucleon. The results unambiguously show that it is
different.

\end{abstract}

 \section{Introduction}

As well known, the deep inelastic scattering (hereafter DIS) of leptons
on nucleons begins by the formation of parton with the size (Compton
wave length in the rest frame) $(mx)^{-1}=\frac{0.21}x\,$fm,
$x=Q^2/(2mq_0)$, $x$ and $q_0$ are the Bjorken scaling variable and the
energy of virtual photon in the rest frame of nucleon. Accordingly
three interaction regions are inherent for the DIS on nuclei:

I. Correlation region, $0<x<0.2$. In this region the size of parton
exceeds the distance between nucleons $r_0=1.2\,$fm, and therefore two
or even several nucleons take part in the process. For this reason the
correlations between nucleons, both short-range and long-range ones,
are of importance.

II. One-nucleon region, $0.2<x<0.8$. In this region $(mx)^{-1}<r_0$,
and therefore the virtual $\gamma$-quantum is absorbed by one nucleon
only.

III. Competition region, $0.8<x\sim1$. In this region for a not very
large $Q^2$ the competition occurs between DIS, elastic lepton--nucleon
scattering  and the possible formation of heavier baryons through the
reaction $\ell N\to\ell'B$, $B$ being $\Delta_{33}$, $N^*$ etc.

In the one-nucleon region we are dealing with the in-medium nucleon
structure function; $F_{2m}(x,Q^2,\bmp,\varepsilon)$ depending upon
the momentum \bmp\ and binding energy $\varepsilon$ of nucleon in
nucleus in addition to $x$ and $Q^2$ must be averaged over the
energy-momentum distribution $S(\bmp,\varepsilon)$ of nucleon, {\em
i.e.}

\begin{equation}
\frac{F_{2A}(x,Q^2)}{x_A}\ =\ \int d^3pd\varepsilon
S(\bmp,\varepsilon)
\,\frac{F_{2m}(x,Q^2,\bmp,\varepsilon)}{x_N}\ ,
\end{equation}
where $x_A=\frac{Amx}{M_A}$ is the scaling variable of target nucleus,
$M_A$ is its mass and $x_N$ is the in-medium scaling variable of
nucleon. The immediate question is as follows: is the in-medium
structure function $F_{2m}$ the same as that of free nucleon? Note that
in Eq.~(1) we used the in-medium scaling variable $x_N$ which is
different from $x$. The usual hope was that choosing an appropriate
definition of $x_N$ one may absorb the in-medium dependence of the
function $F_{2m}$ and describe the data using the free-nucleon
structure function, {\em i.e.} putting $F_{2m}(x,Q^2,\bmp,\varepsilon)
=F_2(x_N,Q^2)$. The analysis of the available data showed that this is
not the case \cite{1}. But as discussed in \cite{2} all the previous
calculations are based on seemingly evident but erroneous assumption
that the quantity $S(\bmp,\varepsilon)$ is the ground-state spectral
function of the target nucleus. Actually it is the spectral function of
the doorway states for one-nucleon transfer reactions. Indeed, the
nucleon hole (which is just the relevant doorway state) is formed in
the ground state of target nucleus when the struck nucleon is destroyed
by DIS. This state is not the eigenstate of nuclear Hamiltonian thus
being fragmented over the actual states of residual nucleus because of
the correlations between nucleons. The observed spreading width of the
hole states is 20~MeV \cite{3} the fragmentation time thus being
$3\cdot10^{-23}$sec. But the interaction times of DIS is
$2q/Q^2=(mx)^{-1}=\frac{0.7}x\cdot10^{-24}$sec thus being less than
$3\cdot10^{-24}$sec for $x>0.3$. So the DIS interaction time in the
one-nucleon region is an order of magnitude less than that of the
fragmentation and therefore the correlation processes do not have time
to come into play. As a consequence the quantity $S(\bmp,\varepsilon$)
entering (1) is the spectral function of the doorway states. As
discussed in \cite{4} it can be unambiguously calculated in a
model-independent way in contrast to the ground-state spectral
function. So the theory of doorway states provides a natural way for
testing the models of nucleon structure functions in nuclei.

In \cite{2} we performed the EMC calculations assuming the nucleon
structure function in the doorway state $\lambda$ to be the same as
that of free nucleon however dependent upon the in-medium scaling
variable in this state:
\begin{equation}
F_{2m}(x,Q^2,\bmp,\varepsilon_2)=F_2(x_N,Q^2), \quad
x_N=\frac{mx}{m+\varepsilon_\lambda-\beta p_3}, \quad
\beta=\frac{|\bmq|}{q_0}=\left(1+\frac{4m^2x^2}{Q^2}\right)^{1/2},
\end{equation}
where $\varepsilon_\lambda<0$ is the nucleon binding energy in the
state $\lambda$ and the axis 3 is chosen along the momentum of
virtual photon. The results do not agree with all the available EMC
data thus indicating that $F_{2m}$ is different from $F_2$. In the
present work we are testing two different choices of the in-medium
scaling variable, the first belonging to Molochkov \cite{5} and the
second to Pandharipande and coworkers \cite{6}.

\section{Analysis}
\subsection{Molochkov's definition of \boldmath $x_N$}

Using the Bethe--Salpeter technique Molockhov derived the following
expression of the nucleon structure function in nucleus:
\begin{equation}
\frac{F_{2A}(x,Q^2)}{x_A}=\int\frac{id^4p}{(2\pi)^4} \frac{F_2(x_N,
Q^2)}{x_N} \frac{|V_{AN}(P_A,p)|^2}{(p^2-m^2)^2
\left((P_A-p)^2-M^2_{A-1}\right)}, \quad x_N=\frac{mx}{p_0-\beta
p_3}\,.
\end{equation}
The vertex $V_{AN}(P_A,p)$ describes the wave function of nucleon in
nucleus, see Eq.~(6). The meaning of other entering quantities is clear
from Fig.~1, where Eq.~(3) is graphically represented. In a more
detailed form

$$
\hspace*{-10cm}
\frac{F_{2A}(x,Q^2)}{x_A}\ =\ \int\frac{id^3pdp_0}{(2\pi)^4}
\eqno{(3a)}
$$
$$
\times\
\frac{|V_{AN}(p_A,p)|^2}{(p_0-e_p+i\delta)^2(p_0-(M_A+E_{A-1})+i\delta)
(p_0-(M_A-E_{A-1})-i\delta)(p_0+e_p-i\delta)^2}.
$$
The integrand has the first-order pole $p_0=M_A-E_{A-1}$ and the
second-order one $p_0=-e_p=-(m^2+\bmp^2)^{1/2}$ (the latter is
negligible because the function $F_2(x,Q^2)$ vanishes at negative $x$
values) in the upper half-plane of $p_0$ and the second-order pole
$p_0=e_p$ together with the first-order one $p_0=M_A+E_{A-1}$ in the
lower half-plane. For the doorway state $\lambda$ of heavy nucleus
\begin{equation}
E_{A-1}\ =\ \left((M_A-m-\varepsilon_\lambda)^2+\bmp\right)^{1/2}\
\cong\ M_A-m-\varepsilon_\lambda
\end{equation}
(in this case the recoil may be neglected) and
$$
E_{A-1}\ =\ e_p
\eqno{(4a)}
$$
for the case of deuteron.

Closing the integration contour over $p_0$ in the upper half-plane we
get $|\delta=e_p-(M_A-E_{A-1})|$
\begin{eqnarray}
F_{2A}(x,Q^2) &=& \frac{Amx}{M_A}\int\frac{d^3p}{(2\pi)^3}
\frac{M_A-E_{A-1}-\beta p_3}{mx}
\nonumber\\
&\times& F_2\left(\frac{mx}{M_R-E_{A-1}-\beta
p_3},Q^2\right)\frac{|V_{AN}(P_A,p)|^2}{2E_{A-1}\delta^2(2e_p-\delta)^2}.
\end{eqnarray}
This is just our result \cite{2} for the in-medium value of the scaling
variable. Comparing Eq.~(5) with Eqs. (29), (30) and (31) of
Ref.~\cite{2} we get
\begin{equation}
\frac{|V_{AN}(P_A,p)|^2}{2(2\pi)^3E_{A-1}\delta^2_\lambda
(2e_p-\delta_\lambda)^2} = \frac{M_A}{\bar
M_A}\,\frac{f_\lambda(p)}{4\pi}\,, \quad
\delta_\lambda=e_p-m-\varepsilon_\lambda\,, \quad \bar
M_A=\sum_\lambda \nu_\lambda(m+\varepsilon_\lambda)
\end{equation}
for the doorway state $\lambda$ ($f_\lambda(p)$ is the nucleon momentum
distribution in this state, see \cite{2} for details) and
$$
\frac{V_{DN}(P_{D,p})}{2(2\pi)^3e_p\delta^2_D(2e_p-\delta_D)^2} =
\frac{M_D}{\bar M_D}\frac{f_D(p)}{4\pi}\,, \quad \delta_D=2e_p-M_D\,,
$$
$$
 \bar M_D=2(M_D-\bar e_p), \quad \bar
e=\int\frac{d^3p}{4\pi}e_pf_D(p)
\eqno{(6a)}
$$
for deuteron.

Molochkov however closed the integration contour in the lower
half-plane so
\begin{eqnarray}
&& \hspace*{-0.7cm}
F_{2A}(x,Q^2)=\frac{Amx}{M_A}\int\frac{d^3p}{(2\pi)^3} \left(
\frac{p_0-\beta p_3}{mx}F_2\left(\frac{mx}{p_0-\beta p_3},Q^2\right)
\right.
\nonumber\\
&&\times\ \left.
\frac{|V_{AN}(P_A,p)|^2}{(p_0+e_p)^2(M_A-E_{A-1}-p_0)(M_A+E_{A-1}-p_0)}
\right)^\prime_{p_0=e_p}
\nonumber\\
&&+\ \frac{Amx}{M_A}\int\frac{d^3p}{(2\pi)^3} \frac{M_A+E_{A-1}-\beta
p_3}{mx}F_2\left(\frac{mx}{M_A+E_{A-1}-\beta p_3},Q^2\right)
\nonumber\\
&&\times\
\frac{|V_{AN}(P_A,p)|^2}{2E_{A-1}(M_A+E_{A-1}-e_p)^2(M_A+E_{A-1}+e_p)^2}.
\end{eqnarray}
Performing the calculations he disregarded the second term of the rhs
and included only part of the derivative in the first term by putting
\begin{eqnarray*}
&& \hspace*{-1.cm}
\left(\frac{p_0-\beta p_3}{mx}F_2\left(\frac{mx}{p_0-\beta
p_3},Q^2\right)\frac{|V_{AN}(P_A,p)|^2}{(p_0+e_p)^2(M_A-E_{A-1}-p_0)
(M_A+E_{A-1}-p_0)}\right)^\prime_{p_0=e_p}
\\
&&=\ \frac{|V_{AN}(P_{A,p})|^2}{4e^2_p(M_A+E_{A-1}-e_p)} \left(
\frac{p_0-\beta p_3}{mx(M_A-E_{A-1}-p_0)}F_2\left(\frac{mx}{p_0-\beta
p_3},Q^2\right)\right)^\prime_{p_0=e_p}.
\end{eqnarray*}
We instead neglected only the $p_0$ dependence of the vertex
$V_{AN}(P,p)$ and included both terms of the rhs of Eq.~(7) thus
obtaining
\begin{eqnarray}
&& \hspace*{-0.7cm}
F_{2A}(x,Q^2)\ =\ \frac{Amx}{M_A}\int \frac{d^3p}{(2\pi)^3}\frac{\left(
1+\frac\delta{e_p}-\frac\delta{M_A+E_{A-1}-e_p}\right)|V_{AN}
(P_{A,p})|^2}{4e^2_p\delta^2(M_A+E_{A-1}-e_p)}
\nonumber\\
&&\times\
\left(\frac{e_p-\Delta-\beta p_3}{mx}F_2\left(\frac{mx}{e_p-\beta p_3},
Q^2\right)+\frac\Delta{e_p-\beta p_3}\dot F_2\left(\frac{mx}{e_p-\beta
p_3},Q^2\right)\right)
\nonumber\\
&&+\ \frac{Amx}{M_A}\int \frac{d^3p}{(2\pi)^3} \frac{|V_{AN}(P_A,p)|^2}{
2E_{A-1}(M_A+E_{A-1}-e_p)^2(M_A+E_{A-1}+e_p)^2}
\nonumber\\
&&\times\
\frac{M_A+E_{A-1}-\beta p_3}{mx}F_2\left(\frac{mx}{M_A+E_{A-1}-\beta
p_3},Q^2\right),
\end{eqnarray}
where
\begin{equation}
\Delta=\left(1+\frac\delta{e_p}-\frac\delta{M_A+E_{A-1}-e_p}
\right)^{-1}\delta, \quad \bigg(~\bigg)^\prime=\frac\partial{\partial
p_0}\bigg(~\bigg), \quad \dot F_2(x,Q^2)=\frac{\partial
F_2(x,Q^2)}{ \partial x}.
\end{equation}

First consider the doorway state $\lambda$ in heavy nucleus. As follows
from Eqs. (5) and (6)
\begin{eqnarray}
&& \hspace*{-0.5cm}
\frac{\left(1+\frac{\delta_\lambda}{e_p}-\frac{\delta_\lambda}{2E_{A-1}
-\delta_\lambda}\right)|V_{AN}(P,p)|^2}{4(2\pi)^3e^2_p(2E_{A-1}-
\delta_\lambda)\delta^2_\lambda}=\frac{M_A}{\bar
M_A}\frac{f_\lambda(p)}{4\pi}\frac{\left(e_p+\delta_\lambda-e_p
\frac{\delta_\lambda}{2E_{A-1}-\delta_\lambda}\right)
(2e_p-\delta_\lambda)^2
E_{A-1}}{4e^3_p E_{A-1}\left(1-\frac{\delta_\lambda}{2E_{A-1}}\right)}
\nonumber\\
&& \cong\ \frac{M_A}{\bar M_A}\frac{f_\lambda(p)}{4\pi}
\frac{4e_p(e_p+\delta_\lambda)\left(e_p-\delta_\lambda
+\frac{\delta^2_\lambda}{4e_p}\right)}{4e^3_p}\cong \frac{M_A}{\bar
M_A}\frac{f_\lambda(p)}{4\pi}\left(1-\frac{\delta^2_\lambda}{e^2_p}
\right)\cong\frac{M_A}{\bar M_A}\frac{f_\lambda(p)}{4\pi}
\end{eqnarray}
because the small quantities $\frac{\delta_\lambda}{2E_{A-1}}\ll1$ and
$\frac{\delta_\lambda^2}{e^2_p}\ll1$ can be neglected in (10). In the
same way
\begin{equation}
\frac{|V_{AN}(P_A,p)|^2}{2(2\pi)^3E_{A-1}\left((M_A+E_{A-1})^2-e^2_p
\right)^2}=\frac{M_A}{\bar M_A}\,\frac{f_\lambda(p)}{4\pi}\left(
\frac{\delta_\lambda(2e_p-\delta_\lambda)}{(M_A+E_{A-1})^2-e^2_p}
\right)^2,
\end{equation}
so
\begin{eqnarray}
&& \hspace*{-1cm}
F_{2A,\lambda}(x,Q^2)=\frac{Amx}{\bar M_A}\int\frac{d^3p}{4\pi}
f_\lambda(p)\left\{\left[\frac{e_p-\Delta_\lambda-\beta p_3}{mx}F_2
\left(\frac{mx}{e_p-\beta p_3},Q^2\right)\right. \right.
\nonumber\\
&&+\ \left.\frac{\Delta_\lambda}{e_p-
\beta p_3}\dot F_2\left(\frac{mx}{e_p-\beta p_3},Q^2\right)\right]
\\
&&+\ \left.
\left(\frac{\delta_\lambda(2e_p-\delta_\lambda)}{(M_A+E_{A-1})^2
-e^2_p}\right)^2 \frac{M_A+E_{A-1}-\beta p_3}{mx}F_2\left(
\frac{mx}{(M_A+E_{A-1}-\beta p_3},Q^2\right)\right\}.
\nonumber
\end{eqnarray}

In the case of deuteron
\begin{eqnarray}
&& \hspace*{-1.cm}
\frac{\left(1+\frac{\delta_D}{e_p}-\frac{\delta_D}{M_D}\right)
|V_{DN}(P_D,p)|^2}{4(2\pi)^3e^2_p\delta_D M_D}=\frac{M_D}{\bar M_D}
\frac{f_D(p)}{4\pi} \frac{(2e_p-\delta_D)\left(1+\frac{\delta_D}{2e_p}
+\frac{\delta_D}{2e_p}-\frac{\delta_D}{M_D}\right)}{2e_p}
\nonumber\\
&&\\
&&=\ \frac{M_D}{\bar M_D}\frac{f_D(p)}{4\pi}\frac{(2e_p-\delta_D)
\left(2e_p+\delta_D-\frac{\delta^2_D}{M_D}\right)}{4e^2_p} \cong
\frac{M_D}{\bar M_D}\frac{f_D}{4\pi}\left(1-\frac{\delta_D^2}{4e^2_p}
\right)\cong\frac{M_D}{\bar M_D}\frac{f_D(p)}{4\pi}
\nonumber
\end{eqnarray}
and
\begin{equation}
\frac{|V_{DN}(P_D,p)|^2}{2(2\pi)^3e_pM^2_D(2e_p+M_D)^2}\ =\
\frac{M_D}{\bar
M_D}\,\frac{f_D(p)}{4\pi}\left(\frac{\delta_D}{2e_p+M_D}\right)^2,
\end{equation}
so
\begin{eqnarray}
&& \hspace*{-1cm}
F_{2D,N}(x,Q^2)=\frac{2mx}{\bar M_D}\int\frac{d^3p}{4\pi}f_D(p)\left\{
\left[\frac{e_p-\Delta_D-\beta p_3}{mx}F_2\left(\frac{mx}{e_p-\beta p_3},
Q^2\right)\right.\right.
\nonumber\\
&&+\ \left.\frac{\Delta_D}{e_p-\beta p_3}\dot F_2\left(
\frac{mx}{e_p-\beta p_3},Q^2\right)\right]
\nonumber\\
&&+\ \left.\left(\frac{\delta_D}{2e_p+M_D}\right)^2
\frac{M_D+e_p-\beta p_3}{mx}F_2\left(\frac{mx}{M_D+e_p-\beta
p_3},Q^2\right)\right\}.
\end{eqnarray}

\subsection{Prescription of Ref. \cite{6}}

This prescription is based on the following consideration: to obtain by
DIS on bound nucleon with the momentum \bmp\ the same hadronic state as
that on free nucleon the momentum transfer \bmq\ must be the same, but
transferred energy $q_0$ must be larger to overcome the binding. As a
result of the energy-momentum conservation
\begin{equation}
e_p+q'\ =\ M_A-E_{A-1}+q_0
\end{equation}
we get
\begin{eqnarray}
&& q'_0=q_0-e_p+(M_A-E_{A-1})=q_0-\delta,
\\
&& Q'^2=Q^2+q^2_0-q'^2_0\cong Q^2+2q_0\delta=
\left(1+\frac\delta{mx}\right)Q^2,
\\
&& x_N=\frac{Q'^2}{2pq'}=\left(1+\frac\delta{mx}\right)
\frac{Q^2}{2(e_pq'_0-\bmp\bmq)} =\frac{mx+\delta}{e_p-\beta p_3},
\end{eqnarray}
so
\begin{equation}
F_{2A}(x,Q^2)=\int\frac{d^3p}{4\pi}\,f(p)\frac{mx}{e_p} \frac{e_p-\beta
p_3}{mx+\delta}\,F_2\left(\frac{mx+\delta}{e_p-\beta p_3},
\left(\frac\delta{mx}\right)Q^2\right).
\end{equation}
It is worth mentioning that putting $\delta=0$ we get the net effect of
the Fermi-motion.

\section{Results}

We calculated the isoscalar part of the EMC ratio~\footnote{The MRST
2002 NLO \cite{mrst} parametrization of the parton distributions in a
free nucleon and the Bonn-B wave function for the deuteron \cite{BB}
were used.}
\begin{equation}
R_A(x,Q^2)\ =\
\frac{F_{2A,p}(x,Q^2)+F_{2A,n}(x,Q^2)}{F_{2D,p}(x,Q^2)+F_{2D,n}(x,Q^2)},
\end{equation}
where the denominator is the structure function of deuteron, and
\begin{equation}
F_{2A,p}(x,Q^2)=\frac1Z\sum^{(p)}_\lambda\nu_\lambda F_{2A,\lambda}
(x,Q^2), \quad F_{2A,n}(x,Q^2)=\frac1N\sum^{(n)}_\lambda F_{2A,\lambda}
(x,Q^2).
\end{equation}

According to the Caushy's theorem the results of the calculations using
our prescription \cite{2}, formula (5), and the Molochkov's one
\cite{5}, , formula (8), must coincide (the difference may be caused by
the approximations /neglection of some small contributions/ used in one
or another method). It is impossible to demonstrate this analytically
because of the absence of analytical expressions for both the structure
functions and the vertices (it is interesting to mention in this
connection that neglecting the difference between $\Delta$ and $\delta$
in formula (8) we get
\begin{eqnarray*}
&& \hspace*{-0.7cm}
\frac{e_p-\delta-\beta p_3}{mx}\,F_2\left(\frac{mx}{e_p-\beta p_3},
Q^2\right)+\frac\delta{e_p-\beta p_3}\,\dot F_2\left(\frac{mx}{e_p-
\beta p_3}, Q^2\right)
\nonumber\\
&&=\ \frac{e_p-\delta-\beta
p_3}{mx}\left(F_2\left(\frac{mx}{e_p-\beta
p_3},Q^2\right)+\left[\frac{mx}{e_p-\delta-\beta p_3}
-\frac{mx}{e_p-\beta p_3}\right]\dot F_2\left(\frac{mx}{e_p-\beta
p_3},Q^2\right)\right)
\end{eqnarray*}
the quantities in the parenthesis thus being the first two terms of the
Tailor's expansion of the function $F_2\left(\frac{mx}{e_p-\delta-\beta
p_3},Q^2\right)$, see (5)~). But the numerical results for the ratios
are found to be the same. The results for the deuteron structure
functions within both the above prescriptions are shown in Table~1. The
comparison clearly shows that the coincidence occurs for the deuteron
structure functions too. It is important to mention that performing the
Molochkov's calculations we neglected the possible $p_0$ dependence of
the nuclear vertices $V_{AN}(P_A,p)$. The coincidence of the
calculations within the methods \cite{2} and \cite{5} indicates that
such a dependence is insignificant.

The results of the calculations using the prescriptions \cite{2,5,6}
for the in-medium scaling variable $x_N$ and that for the Fermi-motion
are shown in Fig.~2 together with the SLAC data \cite{7}. As clearly
seen from the figure none of the prescriptions for $x_N$ leads to
agreement with experiment. The same result is obtained for the EMC data
\cite{8}--\cite{12}. This enables us to state that the structure
function of nucleon in nucleus is different from that of free nucleon.

The $Q^2$ dependence of the results is insignificant. This is
illustrated in Fig.~3 where the EMC ratios for $^{56}$Fe are calculated
at $Q^2=5\rm\,GeV^2$ and $\rm100\,GeV^2$.

\begin{table}
\caption{Comparison between the ratios
$D/np=F_{2D}(x,Q^2)\left(F_{2n}(x,Q^2)+F_{2p}(x,Q^2)\right)^{-1}$
calculated within the prescriptions \cite{2} and \cite{5}.}

\begin{center}
\begin{tabular}{|c|c|c|c|}
\hline
$x$ &  $Q^2$ & $D/np$ [2]  & $D/np$ [5]\\
\hline
&&&\\
  .100  &  5.0   &   .9989  &      .9993\\
  .123   &   5.0  &    .9984  &       .9988\\
  .148   &   5.0  &     .9977 &         .9982\\
  .205   &   5.0    &    .9960    &      .9965 \\
  .235   &   5.0   &     .9948    &      .9954\\
  .268   &   5.0   &     .9935   &       .9941  \\
  .303   &   5.0    &    .9920       &   .9926\\
  .340   &   5.0  &    .9905   &  .9910\\
  .380   &   5.0    &    .9888     &  .9893 \\
  .420   &   5.0  &   .9875 &    .9878\\
  .460   &   5.0     &   .9864  &   .9865\\
  .500   &   5.0  &      .9862  &    .9861\\
  .540   &   5.0  &   .9871    &   .9866 \\
  .580   &   5.0  &   .9900    &   .9888\\
  .620   &   5.0  &      .9961     &   .9942 \\
  .660   &   5.0  &    1.0075   &   1.0045\\
  .700   &   5.0  &   1.0290    &   1.0243\\
  .740   &   5.0  &   1.0656     &    1.0582\\
  .780   &   5.0  &   1.1316  &    1.1196\\
  .820   &   5.0  &   1.2855    &    1.2642\\
  .860   &   5.0    &   1.4979 &    1.4577\\
\hline
\end{tabular}
\end{center}
\end{table}
\clearpage

\begin{figure}
\centerline{
\epsfig{file=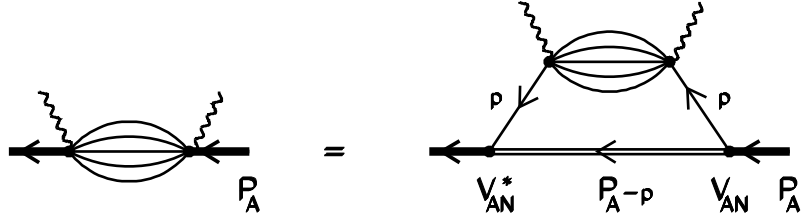,width=12cm}}
\caption{Graphical representation of Eq. (3).}
\end{figure}

\begin{figure}
\centerline{
\vspace{0.2cm}\hspace{0.1cm}
\epsfig{file=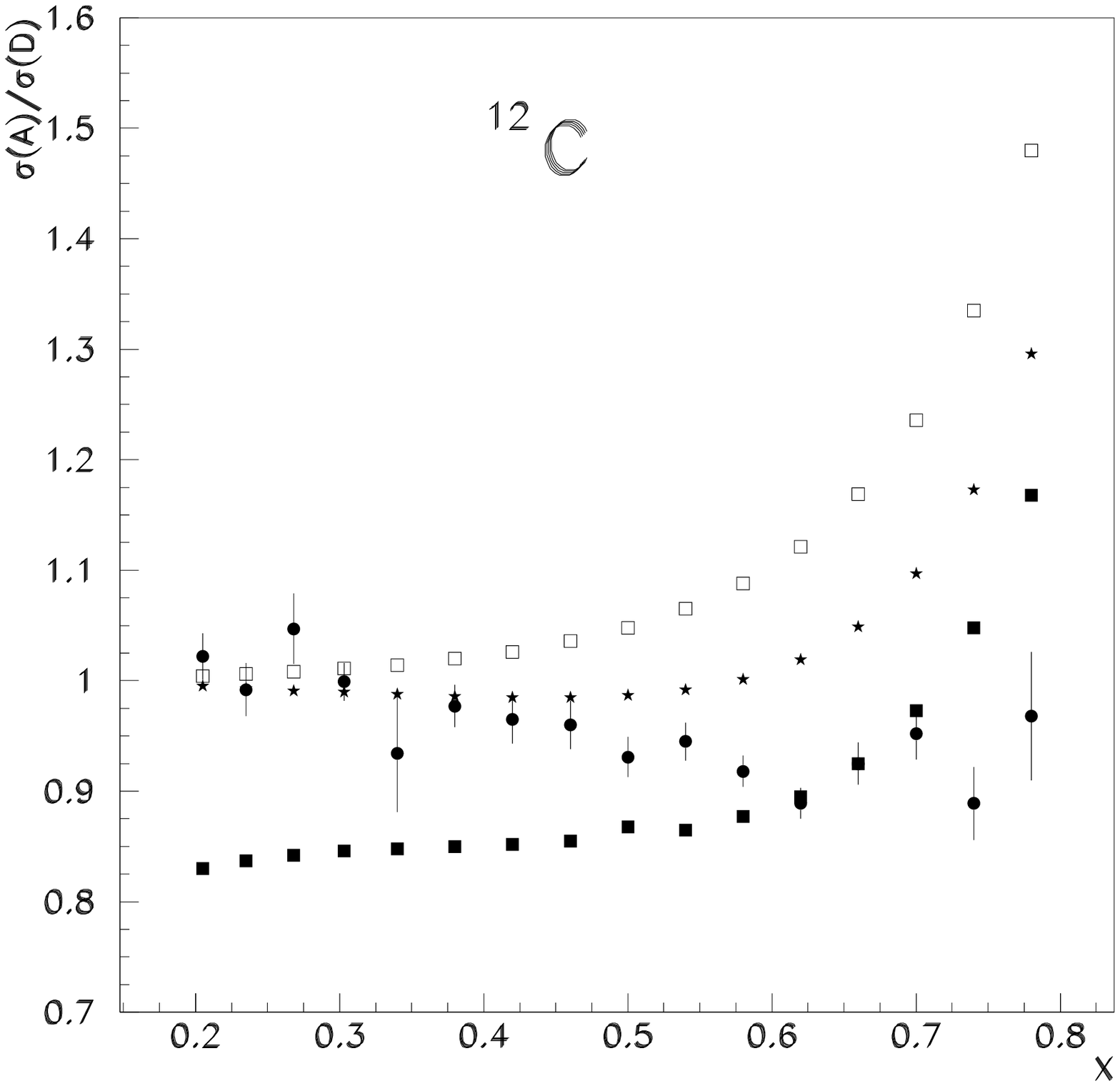,width=7cm}
\epsfig{file=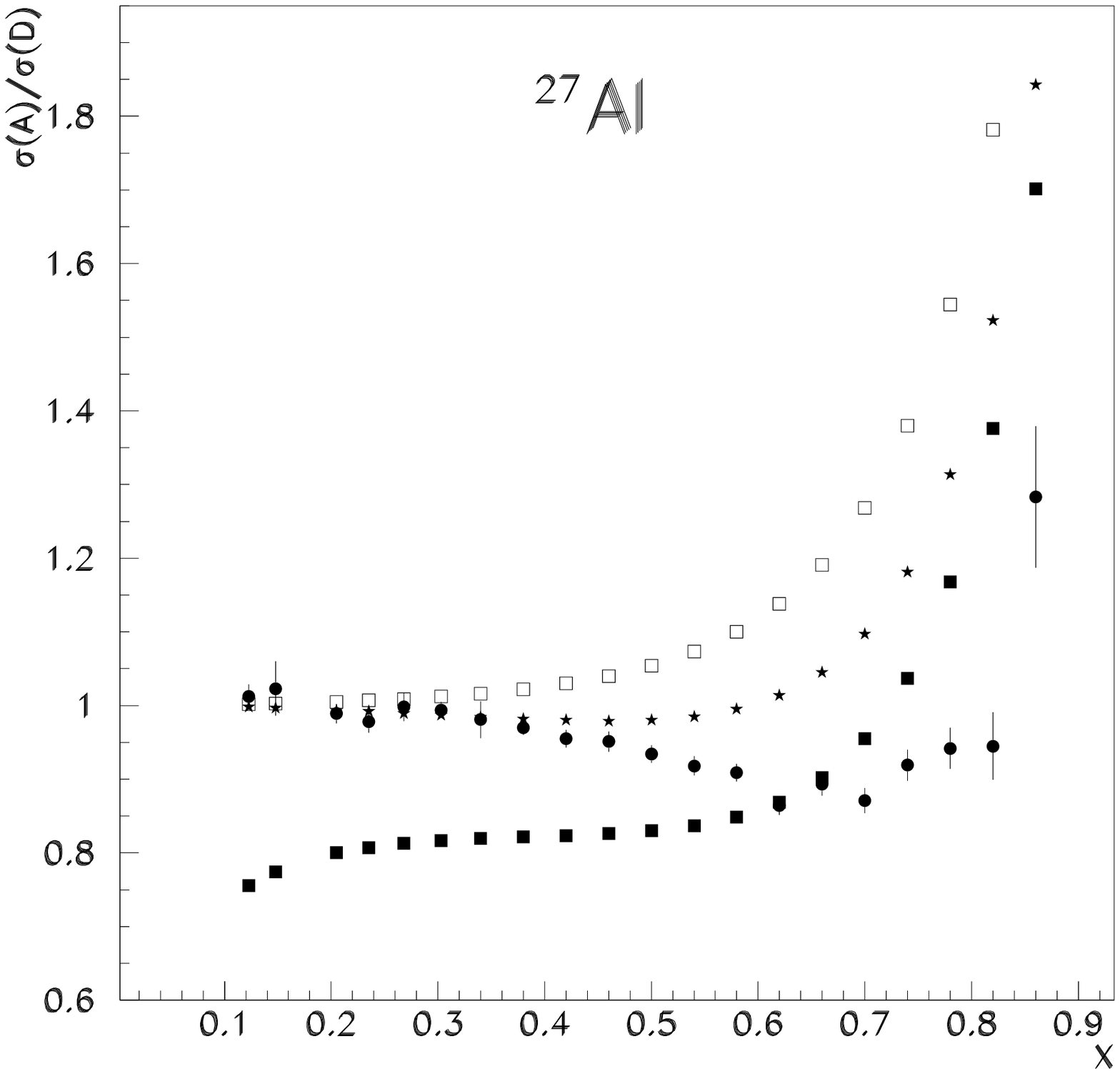,width=7cm}
}
\vspace*{-0.5cm}
\centerline{\vspace{0.2cm}\hspace{0.1cm}
\epsfig{file=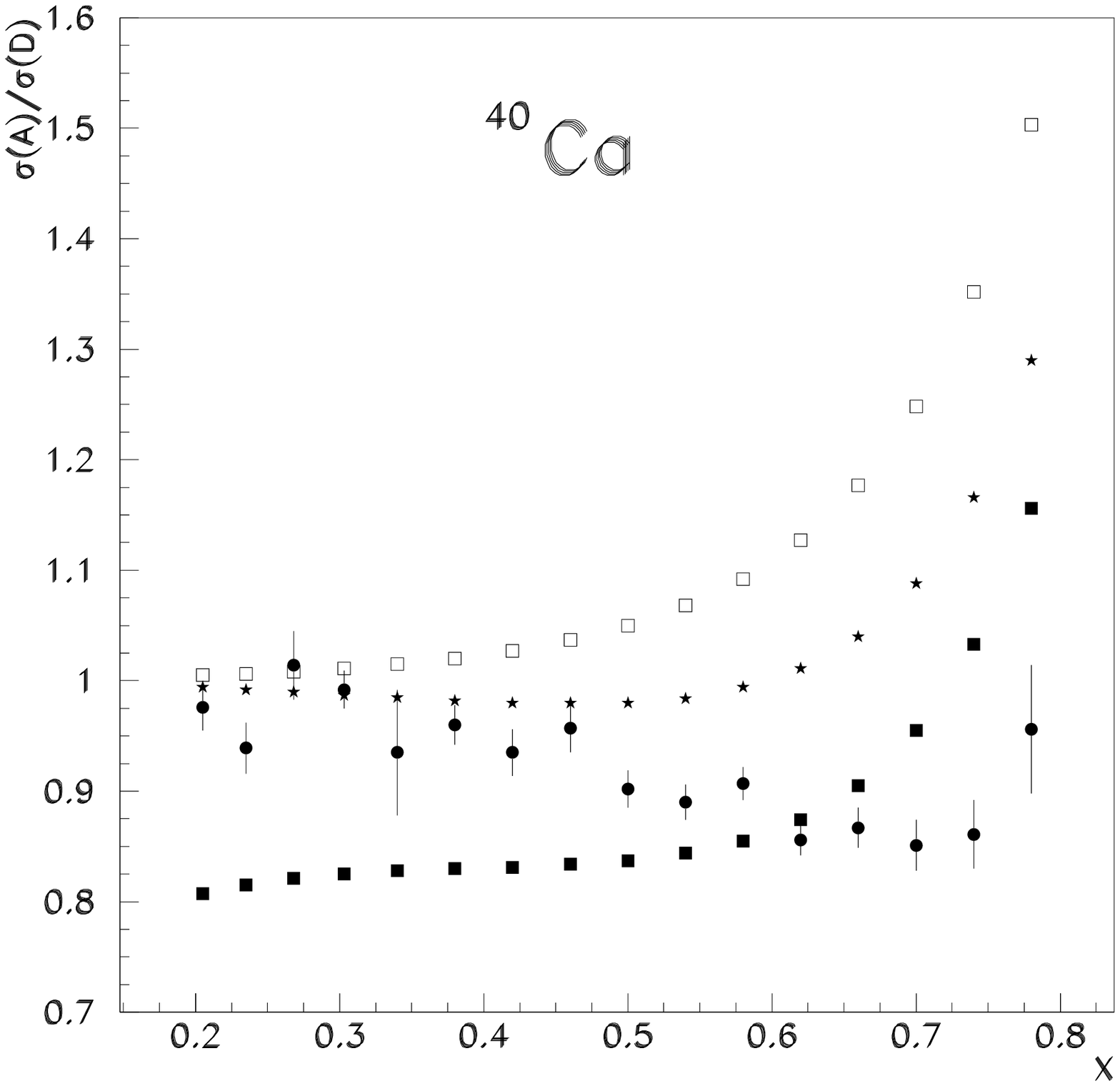,width=7cm}
\epsfig{file=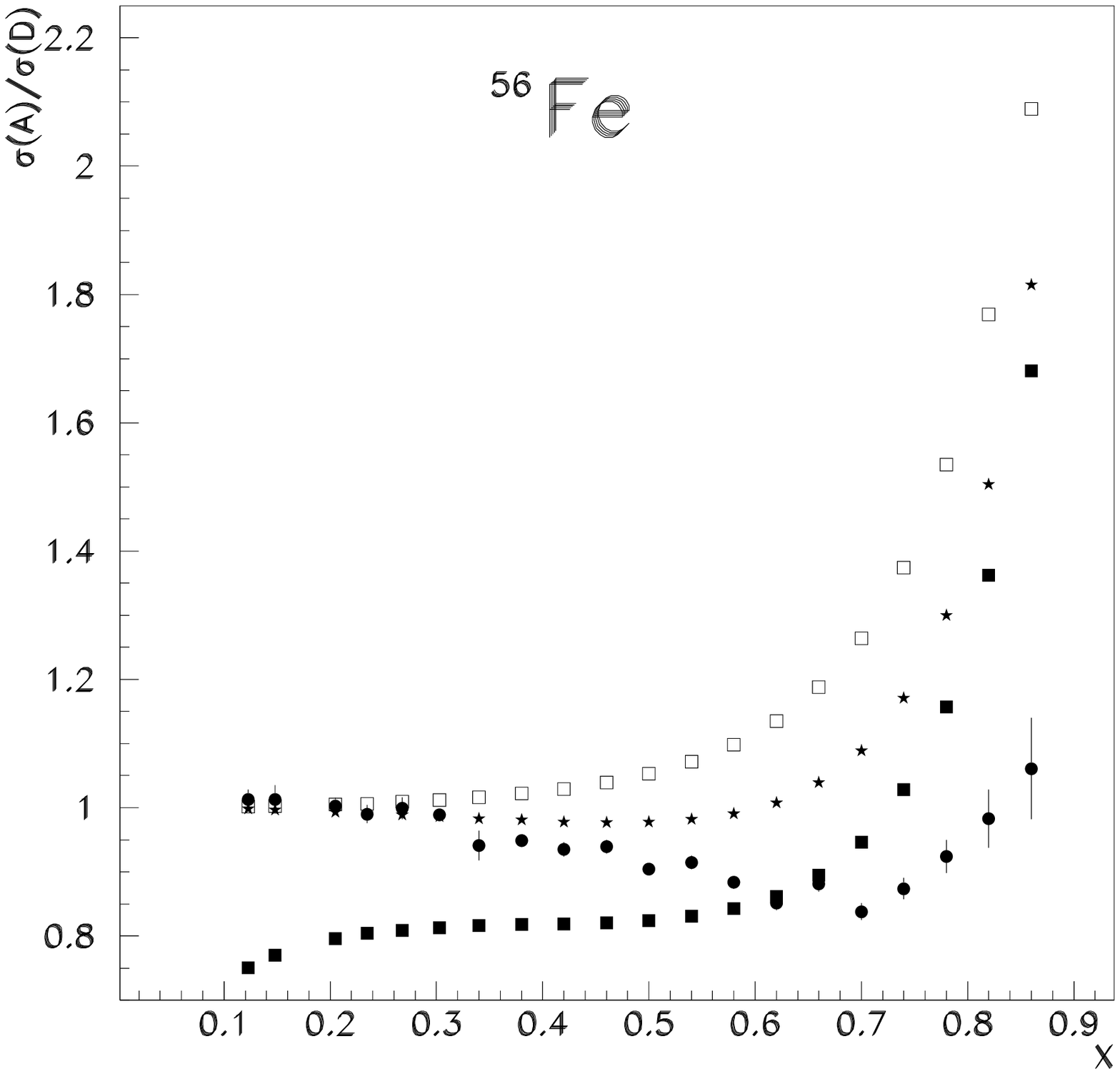,width=7cm}
}
\caption{The EMC ratios at $Q^2=5\rm\,GeV^2$ (left) and $100\rm\,GeV^2$
(right) for $^{56}$Fe; the data are from ref.[9]
 (asterisks; the results \cite{2} and \cite{6} are
undistinguishable), \cite{6} (filled squares) and those for the
Fermi-motion (open squares).}

\end{figure}

\begin{figure}
\centerline{\vspace{0.2cm}\hspace{0.1cm}
\epsfig{file=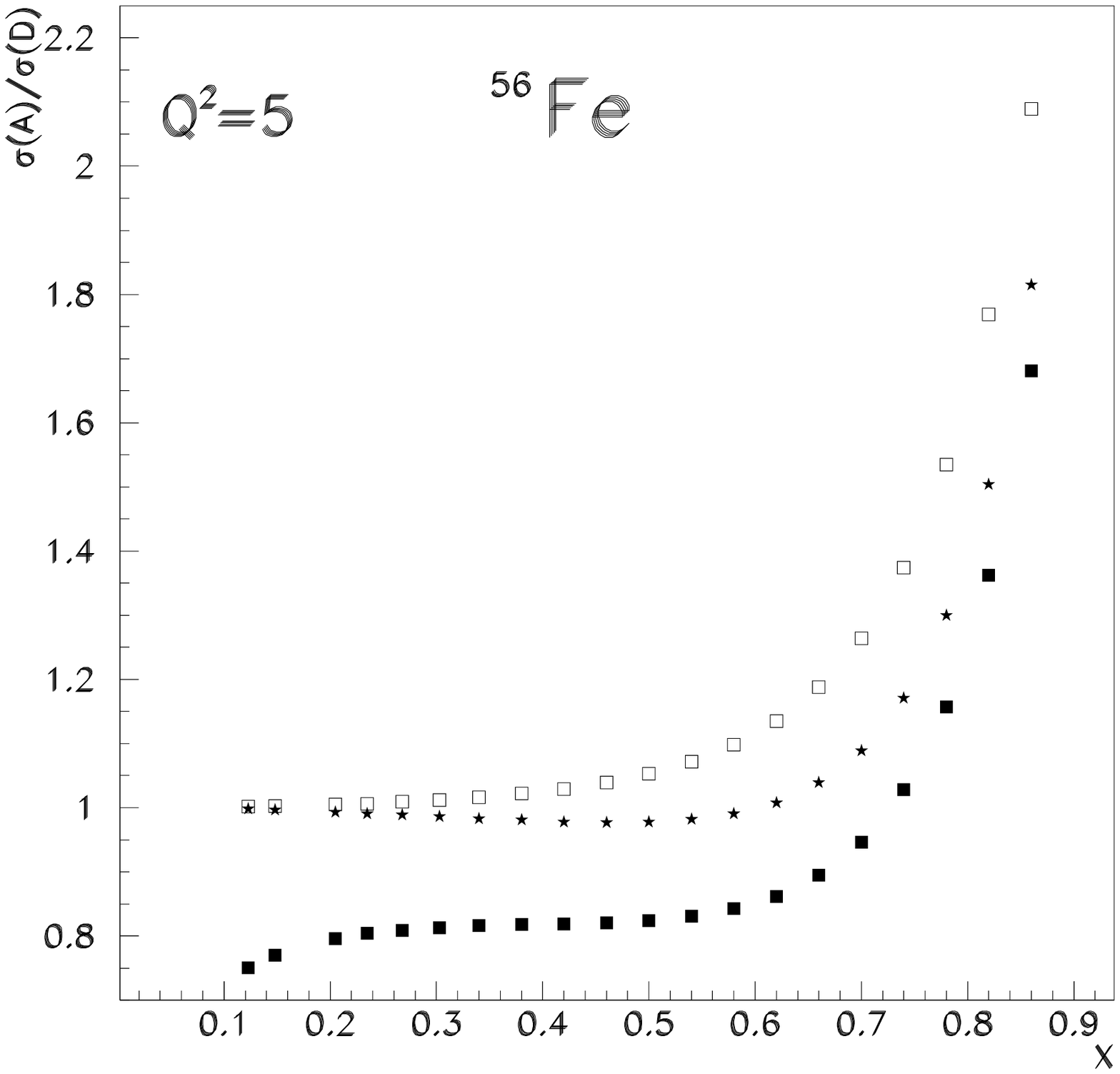,width=7cm}
\epsfig{file=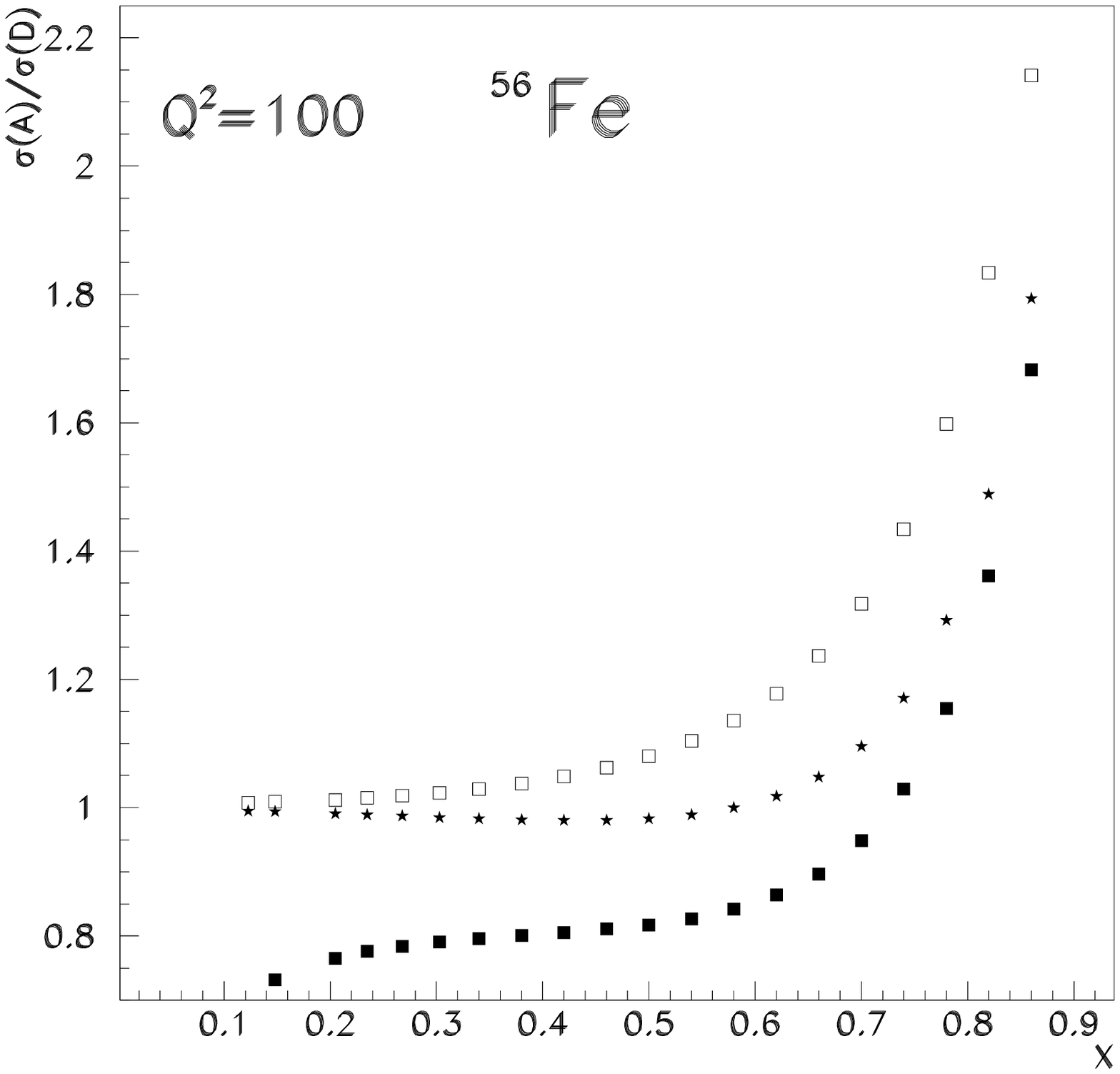,width=7cm}
}
\caption{The EMC ratios at $Q^2=5\rm\,GeV^2$ within the prescriptions
\protect\cite{2,5}
 (asterisks; the results \cite{2} and \cite{6} are
undistinguishable), \cite{6} (filled squares) and those for the
Fermi-motion (open squares).}
\end{figure}

\end{document}